\documentclass{article}
\usepackage{authblk}
\usepackage{color}
\usepackage{graphicx}
\usepackage{amsfonts}
\usepackage{algorithm,algorithmic}
\usepackage{lineno}

\title{A shortest-path based clustering algorithm for joint human-machine analysis of complex datasets}
\author[1,2]{\rm Diego Ulisse Pizzagalli}
\author[2,*]{\rm Santiago Fernandez Gonzalez}
\author[1,*]{\\ Rolf Krause}
\affil[1]{\small Universit\`a della Svizzera Italiana, Faculty of Biomedical Sciences, Institute for Research in Biomedicine, CH6500 Bellinzona}
\affil[2]{\small Universit\`a della Svizzera Italiana, Institute of Computational Science, CH6900 Lugano}
\affil[*]{Correspondence should be addressed to santiago.gonzalez@irb.usi.ch and rolf.krause@usi.ch}

\date{\today}
\begin{document}
\maketitle

\begin{abstract}
	Clustering is a technique for the analysis of datasets obtained by empirical studies in several disciplines with a major application for biomedical research. Essentially, clustering algorithms are executed by machines aiming at finding groups of related points in a dataset. However, the result of grouping depends on both metrics for point-to-point similarity and rules for point-to-group association. Indeed, non-appropriate metrics and rules can lead to undesirable  clustering artifacts. This is especially relevant for datasets, where groups with heterogeneous structures co-exist.
	In this work we propose an algorithm that achieves clustering by exploring the paths between points. This allows both, to evaluate properties of the path (such as gaps, density variations, etc.), and expressing the preference for certain paths. Moreover, our algorithm supports the integration of existing knowledge about admissible and non-admissible clusters by training a path classifier. We demonstrate the accuracy of the proposed method on challenging datasets including points from synthetic shapes in publicly available benchmarks and microscopy data.
\end{abstract}

\section{Introduction}
Clustering algorithms aim at grouping points. This task has critical consequences in pattern discovery and knowledge extraction from complex datasets of unlabeled data, which are commonly produced by empirical studies and data-driven research.
Indeed, identified groups can indicate the presence of multiple phenomena (i.e. populations) and their structure (hereafter, shape) can indicate the relationship between the elements in the groups.
However, being the solution of clustering problem not unique, it is driven towards certain properties by different metrics for point-to-point similarity and rules for point-to-cluster association.
Indeed, metrics and rules generally restrict the set of admissible solutions. As a consequence, when metrics and rules are not chosen appropriately, clustering artifacts are introduced. This is especially relevant when analyzing datasets which have points from populations with arbitrary and heterogeneous structure. For example, common implementations of the K-MEANS algorithm \cite{macqueen1967some} use the Euclidean-distance as metric for point-to-point similarity, while points are associated to the group with the Euclidean-closer centroid. Despite the efficiency of these methods in clustering linearly-separable groups (i.e. globular-like), artifacts can be introduced when separating non-convex shapes. Indeed, if a point should be associated to a cluster having a farther centroid than a second cluster in close proximity, the two clusters will be cut by a non-realistic linear boundary.\\
Interestingly, methods based on density have allowed to overcome these limitations. Arbitrary shapes are reconstructed according with a local density-connectivity criterion. For example, DBSCAN\cite{DBSCAN} considers a point belonging to a cluster, if sufficiently many points in a neighborhood are common (density reachable). As a result, the association rule of DBSCAN correctly identifies clusters with any shape having sufficient density. However, when two or more clusters are present and in close-proximity, a wide density reachability threshold could join them. Moreover, an excessively strict threshold can fail in detecting the clusters.\\
Recently, Rodriguez and Laio \cite{rodriguez2014clustering} proposed a remarkable strategy that achieves Clustering by finding Density Peaks (CDP). CDP, addresses the limitations of DBSCAN by initially finding density peaks and using them to separate clusters. Density peaks are considered as points surrounded by a sufficiently many points with lower density. These neighbors with lower density make density peaks distant. Therefore, CDP identifies density peaks by detecting the outliers of a density-distance plot. This task can be performed either automatically or manually by an user visually inspecting the density-distance plot. Due to the simplicity of this rule CDP can be applied whenever density can be measured which hold in a wide range of applications.\\
Once peaks are identified, each one is assigned to a different cluster with a unique label. Then, CDP processes each remaining point by inheriting the label of the closest point with higher density. Although this rule is applicable to any cluster shape, cases remain where such a local criterion is not optimal.
\\
In this work, we propose to substitute the local association rule of CDP, with the solution of a global optimization problem on a graph which allows to solve the aforementioned problems.
Moreover, expressing these properties as a path cost allows to efficiently propagate a signal from density peaks to the remaining points using the Dijkstra Single Source Shortest Path (SSSP) algorithm\cite{Dijkstra1959} and associating point to the density peaks connected with the shortest path.
Additionally, we demonstrate the possibility to use a trainable path-classifier to define the cost of a path. This allows to adapt the proposed method to specific tasks and datasets using a semi-supervised approach for data-analysis.\\
In conclusion, the proposed data structure and clustering algorithm supports the exploitation of network information\cite{gerber2015improving} and the integration of possibly existing pairwise knowledge, in line with a theory-guided-data-science paradigm\cite{karpatne2017theory}.

\section{Results}

\subsection*{Improving CDP by integrating existing knowledge in path-based clustering}
\begin{figure}[h!]
    \includegraphics[width=\linewidth]{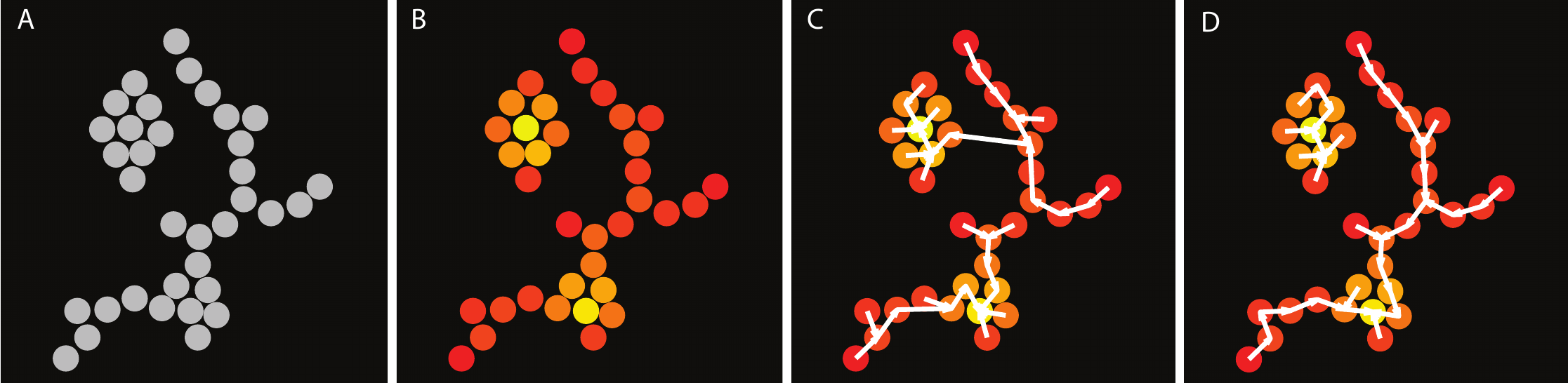}
	\caption{Association rules on a simplified example. \textbf{A.} Raw data-points in a 2d space. \textbf{B.} Color-coded point density (red corresponds to low density, yellow corresponds to high density). \textbf{C.} Local association rule (each point connected to the closest point with a higher density). \textbf{D.} Proposed global association rule using minimax cost function.}
	\label{syntheticdendrites}
\end{figure}

The proposed algorithm retains the strategy of CDP for identifying density peaks but associates each point to the density peak reachable with the optimal path. This allows to consider properties of the path (rather than the difference between two points) for point-to-cluster association.
An example is illustrated in (Figure \ref{syntheticdendrites}, A) where a long-thin projection of a non-globular cluster and a second globular-like cluster are in close proximity. The estimated point-density and density-peaks (Figure \ref{syntheticdendrites}, B) are used by CDP to associate each point the closest point with higher density (Figure \ref{syntheticdendrites}, C - white arrows). In this case, the terminal part of the thin projection is associated to the wrong cluster. By contrast, in (Figure 1, D), properties of the paths have been exploited by the proposed algorithm, to remove the  previously mentioned artifacts. In this case, the preference for multiple small gaps has been expressed using a path cost function that penalizes large gaps in the path, using a minimax path cost function (Figure \ref{syntheticdendrites}, D).

\subsection*{Algorithm}
\begin{figure}[h!]
	\includegraphics[width=\linewidth]{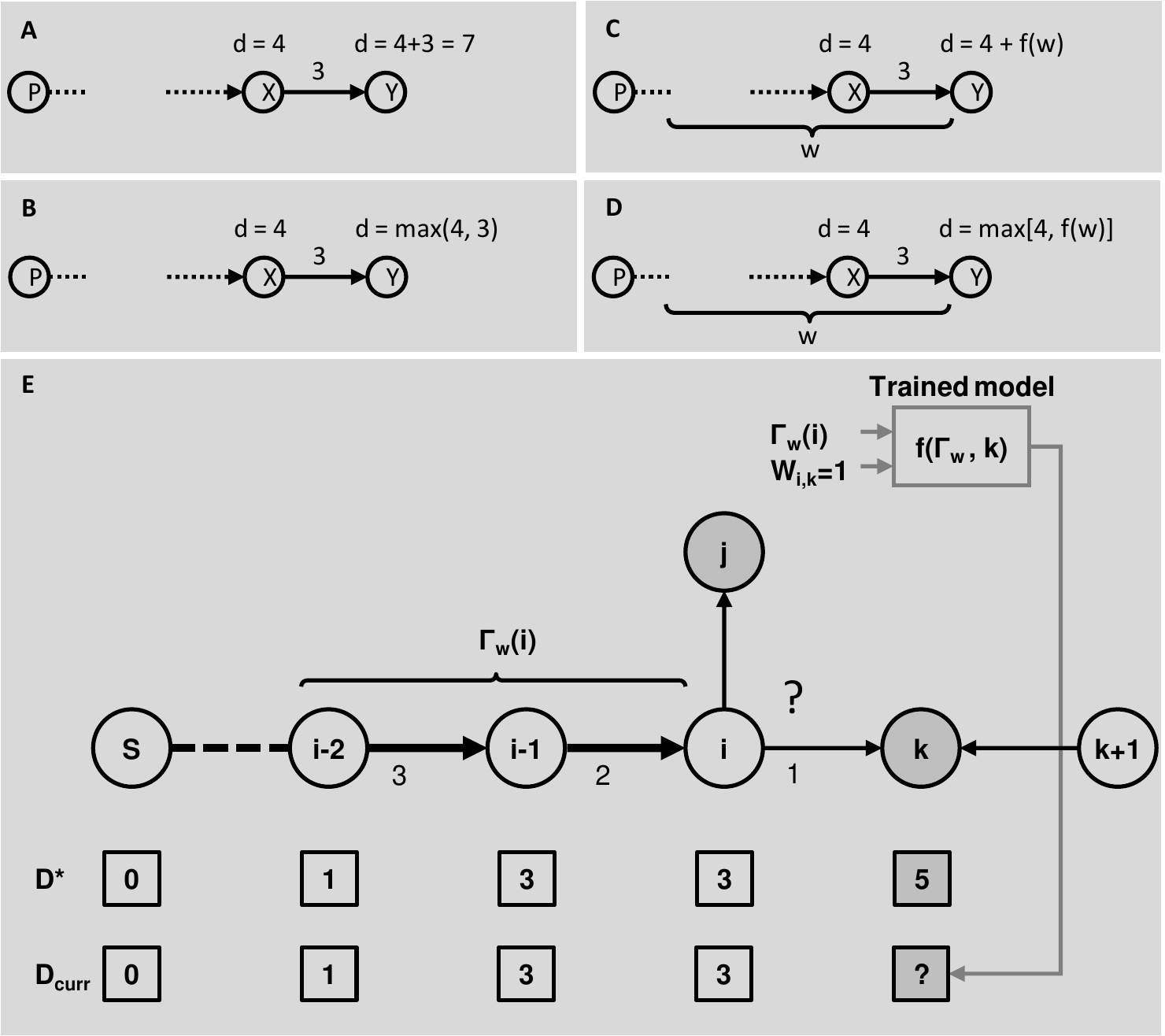}
	\caption{Different non-decreasing path costs. \textbf{A.} Cumulative euclidean cost, used in the standard Dijkstra algorithm. \textbf{B.} $L_\infty$ cost (minimax). This choice avoids gaps and the maximum path cost is bounded by the value of the longest edge in a path. \textbf{C.} Cumulative non-euclidean cost. Here $f(.)$ is the cost of a path defined on a local window $w$. $f(.)$ can derive from a trained classifier to express path preferences \textbf{D.} The non-euclidean path cost $f(.)$ with a minimax formulation. \textbf{E.} Example indicating the execution of the algorithm, using a trained model as path-cost function. A path $\Gamma(i)$ has been identified from the starting node $S$ until to the node $i$. The cost to extending this path to the node $k$ is evaluated as a function of the latest segment of the path and the proposed edge $f(\Gamma_w, k)$. If the proposed path cost $D_{curr}$ is less or equal than the best distance ($D*$), then the edge is included in the current path and the node k is associated to the current cluster.}
	\label{pathcosts}
\end{figure}

The proposed global optimization problem substitutes the local association schema of CDP by computing the Single Source Shortest Path (SSSP) to all the points on a graph. This is achieved by imposing the passage from density peaks and using a customized path-cost function.
For explaining this more in detail, let us make some definitions.
\begin{itemize}
	\item[] Let $G(V,E,w)$ be a weighted graph where a node $v \in V$ is connected to another node $u \neq v$ through an edge $e_{v,u} \in E$ with an associated cost $w_{v,u}: E \to \mathbb{R_+}$.
	\item[] Let $\rho_v$ be the density of point $v$, defined as the number of points in a weighted neighbor according with \cite{rodriguez2014clustering}.
	\item[] Let $\delta_v = \min{\|v-u\|} \quad s.t. \quad \rho_u > \rho_v, \quad (v,u) \in V, $ be the distance of $v$ to the closest point with a higher density.
	\item[] Let $P_\tau \subset V, \quad P_\tau = \{v | v \in V, \quad \rho_v > \tau_\rho \quad \cap \quad \delta_v > \tau_d \}$ be density peaks, i.e. nodes whose local density $\rho_p, p \in P$ is sufficiently high and whose distance $\delta$ to the closest point $q \in V$ with a higher density is greater than a threshold $\tau > 0$.
	\item[] Let $s$ be a fictitious node added to $G$ and connected to each density peak with a negligible cost $\epsilon$
\end{itemize}
The goal is to find the SSSP towards all the points, with $s$ as source, corresponding to the minimal spanning tree having $s$ as root.
Let $\Gamma=\{s, ..., x\}$ be the shortest path connecting the starting node $s$ to an arbitrary point $x$. It is possible to show for a constructive proof (Supplementary material 1) that $x$ is reached, passing through a single density peak $p$ and possibly intermediate points. In this case $ l(c) = l(p) $, i.e. $x$ belongs to the cluster containing $p$.
It is important to notice that, in a fully connected graph, the usage of Euclidean metric for edge cost and the sum of edge costs for path cost, does not solve the problem of associating $x$ to the closest but possibly not optimal peak $p$. However, it is possible to describe the position of $x$ in a cluster according with the path connecting it to a density peak. Hence, properties of the path, such as mediating elements \cite{fischer2003path}, textures\cite{ghidoni2014texture}, density patterns, or user-defined metrics, can be used to associate a meaningful cost. Additionally, this opens the possibility to integrate existing knowledge in a supervised learning approach to prefer certain paths with respect to others by training a path-classifier with examples of admissible and non-admissible paths.
However, to use Dijkstra's SSSP algorithm with a customized path cost function, Beltman optimality must be guaranteed. As a consequence, the path-cost associated to the path $\Gamma' = {s,x_i,x_{i+1}}$ must always be greater or equal than the cost of the sub-path $\Gamma = {s,x_i}$. This restricts the choice to non decreasing path cost functions.
In Figure \ref{pathcosts} are illustrated possible non-decreasing path-cost functions such cumulative euclidean (Figure \ref{pathcosts}, A), minimax (Figure \ref{pathcosts}, A), cumulative with the evaluation of a local path fragment $w$ (Figure \ref{pathcosts}, A) and minimax with the evaluation of a local path fragment $w$.
Amongst these, the minimax formulation defines, for a generic path $\Lambda$ the cost $c(\Lambda)$ as $\max(w_{i,i+1}), {i,i+1} \in \Lambda$. $\Gamma(s,p)$ is "minimax path" if its cost is minimum with respect to all the existing paths from the same source to the same destination. $$\Gamma \quad \textbf{"minimax"} \quad \textbf{if} \quad M(\Gamma) \le M(\Lambda) \quad \forall \Lambda \neq \Gamma$$
The cost of an extended minimax path, which is a minimax path with an additional node, can be computed by a function $f(\Gamma_w, e)$ which computes the cost of following an edge based on properties of the latest $w$ edges and the proposed edge $e$.
In the examples provided in this work we integrated the generic preference for small gaps with respect to large gaps, using a minimax cost function.
In (Figure \ref{pathcosts}, E), the Dijkstra algorithm is applied by computing at runtime the distance between the explored nodes ${S, i-2, i-1, i, k+1}$ and the node $k$. The best distance $D*$ for reaching $k$ from $k+1$ is 5. The proposed update using the current path $\Gamma = {S, ..., i, k}$ has a cost evaluated by a trained path classifier using a minimax formulation.

In terms of computational resources, Dijkstra SSSP algorithm does not require to store the graph on memory, computing the cost for following an edge on demand. Time-wise, the computational complexity of the proposed method remains bounded by Dijkstra's algorithm which is less than quadratic in the number of nodes and can be optimized for sparse graphs.

\subsection*{Segmentation of globular and dendritic cells in confocal microscopy}
\begin{figure}[h!]
	\includegraphics[width=\linewidth]{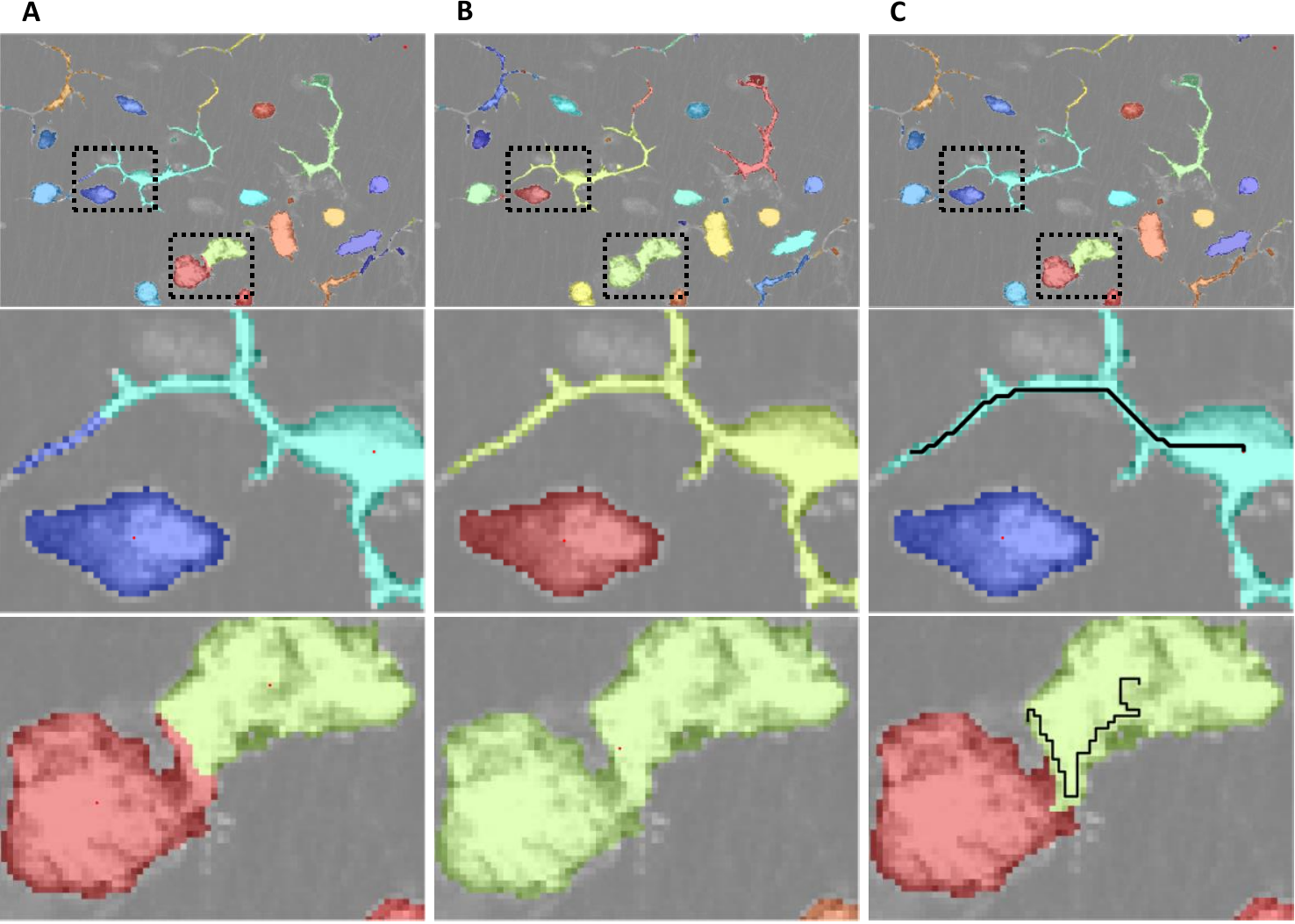}
	\caption{Qualitative results (color-coded labels) produced by different density-based clustering algorithms on the shapes of immune cells. Data were acquired by confocal microscopy and includes murine CD11c+ GFP immune cells in normal conditions. The 2d projection (MIP) is represented. \textbf{A.} CDP \cite{rodriguez2014clustering} using an euclidean metric correctly separates touching cells but associates a piece of dendrite to the wrong cell. \textbf{B.} DBSCAN correctly reconstructs the shape of dendritic cells but is not able to separate touching cells with the same density-reachability criterion. \textbf{C.} Proposed method correctly associating the dendrite of the dendritic cell and separating touching cells. Black lines indicate the optimal path followed by the algorithm, from the cell centroid (density peak) to a point in the dendrite and in the touching region respectively.}
	\label{DCs}
\end{figure}

Microscopy images are considered challenging spatio-temporal biomedical datasets \cite{ulman2017objective,mavska2014benchmark}. This is particularly relevant for acquisitions of immune cells which exhibit high plasticity, non-globular shapes and stochastic contacts \cite{Pizzagalli2018, Beltman2009}. As a consequence, manual cell detection, segmentation and tracking are required, lowering the usability of bio-imaging software \cite{carpenter2012call} and compromising reproducibility. Here we apply our clustering algorithm for grouping pixels. We assume that other image processing tasks such as background-extraction and color co-localization of cells of interest have been already executed with a specific software such as Ilastik \cite{ilastik}, Trainable Weka Segmentation\cite{weka} or Imaris (Bitplane). Qualitative results are presented in (Figure \ref{DCs}), where we applied our algorithm to cluster pixels from a confocal microscopy acquisition of fluorescent-labeled immune cells, including dendritic cells (non-globular) and lymphocyte-like cells (globular).
In (Figure \ref{DCs}, A) the local association rule of CDP yielded to the wrong association of a thin dendrite of a dendritic but correctly allowed to separate touching cells. In (Figure \ref{DCs}, B) DBSCAN correctly reconstructed the shape of dendritic cells but did not separate touching cells. In (Figure \ref{DCs}, C) our algorithm, penalizing large gaps in the paths allowed to connect the pixels to the correct density peaks in the two magnified examples. However, some cases remain where our algorithm splits the cells in multiple parts. To merge these sub-clusters method, a different metric can be used for computing the $\delta$ measure for each peak or methods for merging multiple clusters such as \cite{Mehmood2017} can be applied for post-processing.

\subsection*{Quantitative benchmarking on synthetic datasets}
\begin{figure}[h!]
	\includegraphics[width=\linewidth]{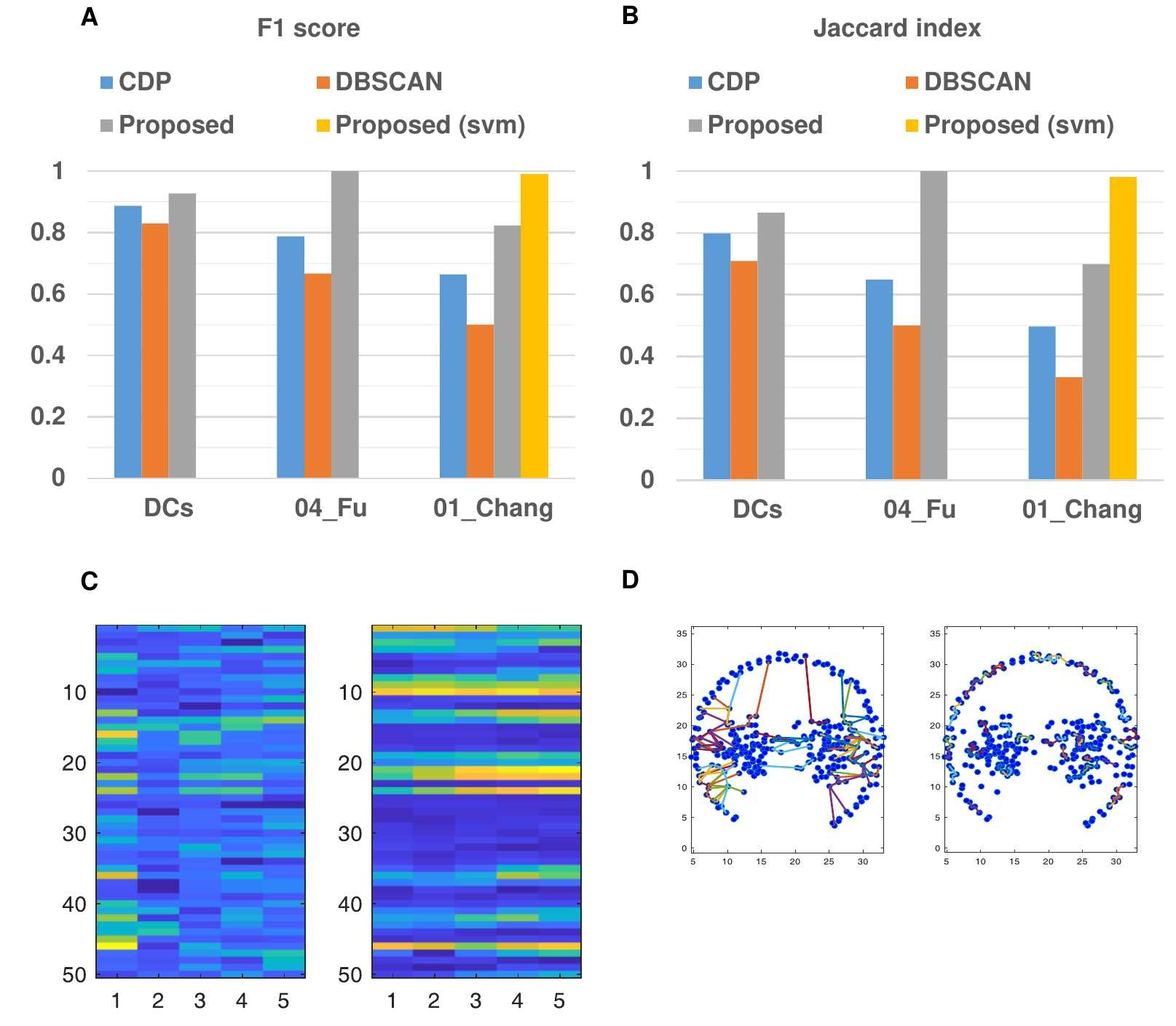}
	\caption{Quantitative results produced by different density-based clustering algorithms CDP\cite{rodriguez2014clustering}, DBSCAN\cite{DBSCAN}, Proposed using a minimax path-cost function, Proposed using a trained model (Support Vector Machine). F1 score (A) is computed vs. the ground truth as $F1 = 2*\frac{Recall * Precision}{(Recall + Precision)}$, the Jaccard index (B) is computed vs. the ground truth as $\frac{TP}{TP+FN+FP}$, where TP are the True positives, Fn the False Negatives, FP the False Positives and TN the True Negatives. For the dataset 01\_Chang a predictive model was trained on 50 desired paths and on 50 undesired paths (D) Which were randomly generated by the script in (Supplementary script 2). These paths were defined over a local window of 5 nodes using the density profile (i.e. an ordered vector of densities) as feature to describe the path (C). Using this constraint the proposed method achieved a F1 score $ge$ 0.99 and a Jaccard index $ge$ 0.98 on the difficult \cite{wiwie2015comparing} Chang\_01 example.}
	\label{quantitiative}
\end{figure}

The \texttt{F1-score} (Figure \ref{quantitiative}, A) and the \texttt{Jaccard index} (Figure \ref{quantitiative}, B) were computed on the synthetic datasets in the a publicly available clustering benchmark ClustEval \cite{wiwie2015comparing}. Moreover, the algorithm was benchmarked against the results produced manually by an operator on the data presented in (Figure, \ref{DCs}).
The proposed method using a general-purpose minimax path-cost function demonstrated a performance increase w.r.t CDP when heterogeneous populations co-exist.
A further increase in performance was obtained by using a Support Vector Machine for path-classification on the challenging synthetic dataset \texttt{01\_Chang}. 50 fragments of good and 50 fragments of bad paths composed by 5 points were described using the point-density as features (Figure \ref{quantitiative}, C).
In comparison to CDP, our method allowed to correctly cluster the outermost elongated structure in (Figure \ref{quantitiative}, D) while still keeping separated the two central clusters in close proximity.

\subsection*{Centroid tracking as a clustering problem}
\begin{figure}[h!]
	\includegraphics[width=\linewidth]{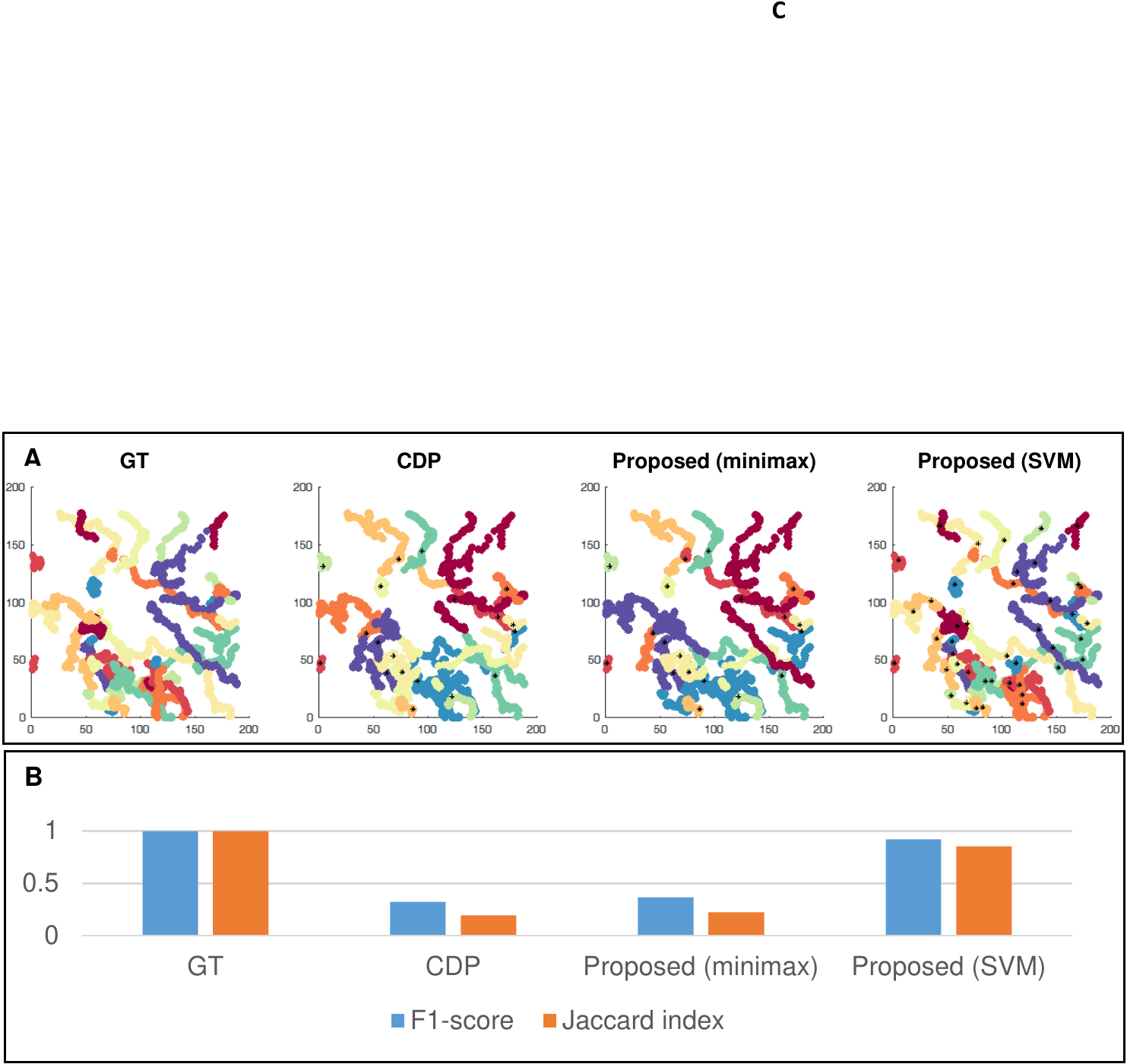}
	\caption{Tracking as a clustering problem. \textbf{A.} Quantitative results with respect to the ground truth (GT) produced by different density-based clustering algorithms CDP, proposed using a minimax path-cost function, proposed using a trained model (Support Vector Machine). \textbf{B.} F1 score and Jaccard Index.}
	\label{4d}
\end{figure}

Considering grouping of points in space-time we demonstrated the applicability of our algorithm for tracking the position of the centroids of immune cells from 4d data (sequences of 3d images). 
Analyzing the migration of immune cells captured in microscopy videos is essential for studying the mechanisms of the immune system \cite{Beltman2009, stein2017dynamic}. The task solved by our algorithm corresponds to the identification of tracks by assigning a track identifier to each centroid. Thanks to the availability of a publicly available dataset of tracks of immune cells, which can serve as ground truth \cite{Pizzagalli2018}, we qualitatively and quantitatively evaluated the performances of our method on the challenging case \texttt{LTDB016\_a} which includes multiple interacting cells.
Without specifying any association rule or spatio-temporal constraint, a Support Vector Machine classifier was trained from the ground truth to recognize admissible and non-admissible track fragments (50 samples of 5 time points for each of the two cases).
The qualitative results are presented in (Figure \ref{4d}, A) with respect to the ground truth (GT), of CDP, our method without the path classifier (using a minimax cost function) and with both the SVM classifier the non-decreasing minimax cost function (SVM). The quantitative results in (Figure \ref{4d}, B) exhibit a F1-score of 33\% for CDP, 37\% of our method without the path classifier and 92\% using the path classifier.
It is important to note that in this application, peaks were assigned picking one single point from each single track from the ground truth. Moreover, here we do not consider RAW images data but we focus on the grouping of a set of centroids,which were manually annotated.

\section{Methods}
\subsection{Data generation}
Immunofluorescence confocal microscopy was performed using a Leica TCS SP5  microscope, with sequential acquisition to limit signal crosstalking. For data shown in Figure 3, a murine trachea was collected and fixed in 4\% paraformaldehyde at 4°C for 1 h. Tracheas were cut in two halves along the long axis. Pieces were placed on a microscopy slide and embedded in Fluoromount™ Aqueous Mounting Medium (Sigma Aldrich).
3D images were acquired using a HCX PL APO CS 20X/0.7 oil immersion objective, z-step was 1 $\mu$m for a total depth ranging from 50 $\mu$m to 112 $\mu$m.
The bioimaging software Bitplane Imaris (v 9.2.1) was used for computing 2D maximum intensity projections (MIP) and for manual cell segmentation. The same software (Coloc functionality) has been used to exclude the background and creating a channel containing only the points belonging to Cd11c-GFP+ cells (Supplementary data 1).

\subsection{Implementation}
The proposed algorithm was implemented in Matlab r2017b. Vectorization and a cache-friendly memory access were considered to optimize the execution of the algorithm which outperforms of the \texttt{graphshortestpath()} routine distributed with the Bioinformatics toolbox of Matlab.
The following libraries have been used:
\begin{itemize}
	\item {DensityClust.m - Implementation of the CDP clustering algorithm\\
	\texttt{fileexchange/53922-densityclust} }
	\item {dbscan.m - Implementation of the DBSCAN clustering algorithm\\
	\texttt{https://www.peterkovesi.com/matlabfns/Misc/dbscan.m}}
	\item {FitCSVM - Implementation of a Supported Vector Machine classifier\\
	\texttt{Matlab bioinformatics Toolbox}}
\end{itemize}
Code and datasets will be deposited and made available along with the publication of the article. Inquiries can be addressed to the corresponding authors.

\subsection{Path classifier}
In the example provided in (Figure \ref{pathcosts}, D) a Support Vector Machine with RBF Kernel was trained on 50 desired paths and 50 non-desired path fragments. Training data were generated using the Matlab code provided in (Supplementary data 1).

\section{Conclusions}\label{conclusions}
The proposed method inherits the advantages of density-based clustering methods but, using a global association rule, is able to overcome the limitations of local connectivity rules. The proposed path-based formulation supports the expression of pat-cost by evaluating path-properties. The non-descreasing minimax formulation allows to solve the proposed clustering problem using the Dijkstra Single Source Shortest Path algorithm which can be efficiently executed on modern parallel computing architectures\cite{harish2007accelerating}, does not require to \textit{a priori} store on memory the adjacency matrix of the graph and doesn't accumulate cost with long paths, extending the compatibility to resource-constrained computes such as ubiquitous devices. Furthermore, the regular structure of images can be exploited to implement optimized access to memory in a cache-friendly way.
\\
Moreover, a non-decreasing path cost supports the integration of existing knowledge by defining an appropriate path-cost function, which can be obtained, for instance via supervised machine learning by training a path classifier.
\\
The application of the proposed method requires the initial execution of CDP to detect density peaks. CDP has a broad applicability and allows to find density peaks in a remarkable simple way, therefore it may limit the application to datasets where density-peaks do not exist.

\section*{Acknowledgements}
We are thankful to Davide Eynard and Benedikt Thelen, Ilaria Arini and Liudmila Karagyaur for technical discussion and support, Rocco D’Antuono, Diego Morone and the IRB imaging facility for microscopy assistance, Tommaso Virgilio and Miguel Palomino-Segura for having generated microscopy data.
This project was supported financially by the Swiss National Science Foundation with the Systemsx.ch grant 2013/124 to D.U.P., by the Swiss National Foundation (SNF) with the grant Ambizione 148183 to S.F.G., by the Marie Curie Reintegration Grant 612742 to S.F.G and by the Center for Computational Medicine in Cardiology (CCMC) at ICS.

\section*{Author contributions}
D.U.P. designed and implemented the algorithm, wrote the manuscript.
S.F.G. supervised the biological aspects of the project, provided microscopy data, wrote the manuscript.
R.K. supervised the computational aspects of the project and wrote the manuscript.

\pagebreak
\section*{Figures}\label{Figures}

\clearpage
\begin{thebibliography}{10}
	
	\bibitem{macqueen1967some}
	J.~MacQueen, {\it et~al.\/}, {\it Proceedings of the fifth Berkeley symposium
		on mathematical statistics and probability\/} (Oakland, CA, USA., 1967),
	vol.~1, pp. 281--297.
	
	\bibitem{DBSCAN}
	M.~Ester, H.-P. Kriegel, J.~Sander, X.~Xu, Others, {\it Kdd\/} (1996), vol.~96,
	pp. 226--231.
	
	\bibitem{rodriguez2014clustering}
	A.~Rodriguez, A.~Laio, {\it Science\/} {\bf 344}, 1492 (2014).
	
	\bibitem{Dijkstra1959}
	E.~W. Dijkstra, {\it Numerische Mathematik\/} {\bf 1}, 269 (1959).
	
	\bibitem{gerber2015improving}
	S.~Gerber, I.~Horenko, {\it Science advances\/} {\bf 1}, e1500163 (2015).
	
	\bibitem{karpatne2017theory}
	A.~Karpatne, {\it et~al.\/}, {\it IEEE Transactions on Knowledge and Data
		Engineering\/} {\bf 29}, 2318 (2017).
	
	\bibitem{fischer2003path}
	B.~Fischer, J.~M. Buhmann, {\it IEEE Transactions on Pattern Analysis and
		Machine Intelligence\/} {\bf 25}, 513 (2003).
	
	\bibitem{ghidoni2014texture}
	S.~Ghidoni, L.~Nanni, S.~Brahnam, E.~Menegatti, {\it Studies in health
		technology and informatics\/} (IOS, 2014), pp. 74--82.
	
	\bibitem{ulman2017objective}
	V.~V. Ulman, {\it et~al.\/}, {\it Nature methods\/} {\bf 14}, 1141 (2017).
	
	\bibitem{mavska2014benchmark}
	M.~{Ma{\v{s}}ka}, {\it et~al.\/}, {\it Bioinformatics\/} {\bf 30}, 1609 (2014).
	
	\bibitem{Pizzagalli2018}
	D.~U. Pizzagalli, {\it et~al.\/}, {\it Scientific Data\/} {\bf 5}, 1 (2018).
	
	\bibitem{Beltman2009}
	J.~B. Beltman, {\it et~al.\/}, {\it Nature reviews. Immunology\/} {\bf 9}, 789
	(2009).
	
	\bibitem{carpenter2012call}
	A.~E. Carpenter, L.~Kamentsky, K.~W. Eliceiri, {\it Nature methods\/} {\bf 9},
	666 (2012).
	
	\bibitem{ilastik}
	C.~Sommer, C.~Straehle, U.~K{\"o}the, F.~A. Hamprecht, {\it 2011 IEEE
		international symposium on biomedical imaging: From nano to macro\/} (IEEE,
	2011), pp. 230--233.
	
	\bibitem{weka}
	I.~Arganda-Carreras, {\it et~al.\/}, {\it Bioinformatics\/} {\bf 33}, 2424
	(2017).
	
	\bibitem{Mehmood2017}
	R.~Mehmood, S.~El-Ashram, R.~Bie, H.~Dawood, A.~Kos, {\it Nature Publishing
		Group\/}  (2017).
	
	\bibitem{wiwie2015comparing}
	C.~Wiwie, J.~Baumbach, R.~R{\"o}ttger, {\it Nature methods\/} {\bf 12}, 1033
	(2015).
	
	\bibitem{stein2017dynamic}
	J.~V. Stein, S.~F. Gonzalez, {\it Journal of Allergy and Clinical Immunology\/}
	{\bf 139}, 12 (2017).
	
	\bibitem{harish2007accelerating}
	P.~Harish, P.~Narayanan, {\it HiPC\/} (Springer, 2007), vol.~7, pp. 197--208.
	
\end{thebibliography}
\end{document}